\documentclass[12pt]{iopart}

\usepackage[latin1]{inputenc}
\usepackage{graphicx,epsfig,color}
\usepackage{wrapfig,rotating}

\begin{document}

\title{Measurements of high $p_{T}$ identified particles $v_{2}$ and $v_{4}$ in $\sqrt{s_{NN}} = 200$ GeV Au+Au collisions by PHENIX}

\author{S. Huang, for the PHENIX Collaboration}

\address{Department of Physics and Astronomy, Station B 1807, Vanderbilt University,
Nashville, TN 37235, USA}

\ead{shengli.huang@vanderbilt.edu}

\begin{abstract}
The $v_{2}$ and $v_{4}$ of pions, kaons and protons  have been
measured by PHENIX in 200 GeV Au+Au collisions up to
$p_{T}$$\sim$6 GeV/c and 4 GeV/c, respectively. The $v_{4}$ of all
these identified particles have been found to scale with the
number of constituent quarks and all these particles have a
similar $v_{4}$/$v_{2}^{2}$ ratio which is close to 0.9. The
scaling behavior of $v_2$ is studied at high $p_{T}$ and a
deviation from the universal scaling is observed for transverse
kinetic energy ($KE_{T}/n_{q}$) higher than 1 GeV.
\end{abstract}


\section{Introduction}
A hot, dense non-hadronic matter has been created at RHIC in
ultra-relativistic heavy ion
collision{\cite{whitepaper}}{\cite{photon}}. The anisotropic flow
coefficients $v_{2}$ and $v_{4}$ provide sensitive information
about the properties of the matter in the earliest stages of the
heavy-ion collisions. The $v_{2}$ of identified hadrons has been
found to obey empirical scaling with the number of constituent
quarks (NCQ) for $KE_{T}$ and provides evidence that partonic
degrees of freedom determine the early dynamics of the
system{\cite{flow}}. In this work, the measurement of $v_{4}$ will
be used to further test this scaling. The $v_{4}$/$v_{2}^{2}$
ratio has been proposed as a probe of ideal hydrodynamics and
related to the degree of thermalization{\cite{thermalization}} in
the system. Accurate measurements of identified particles
$v_{4}$/$v_{2}^{2}$ ratio will constrain the model calculations.
The measurement of high $p_{T}$ identified hadron $v_{2}$ will
allow us to test the limits of the NCQ scaling. If the anisotropic
emission of particles in the high $p_{T}$ region is dominated by
parton energy loss, the NCQ scaling is expected to break since the
energy loss mechanism affects all particle species
similarly{\cite{jet}}{\cite{Hwa}}. Determining the breaking point
in the NCQ scaling will provide information on the limits of
applicability of the hydrodynamic description of the system
dynamics.

\section{Analysis Method}
During Run 7 of RHIC, the PHENIX experiment recorded 5.5 B
minimum-bias 200 GeV Au $+$ Au collision events and 2.2 B have
been used for this work. Two new subsystem detectors were
installed prior to Run 7, which significantly enhanced the PHENIX
capabilities for identified particle anisotropic flow measurements
. A time of flight detector(TOFw) was installed in the west arm of
the PHENIX spectrometer. With $\sigma_{t} = 75$ ps intrinsic
timing resolution, the TOFw detector allows pion/kaon separation
up to $p_{T}$$\sim$2.8 GeV/$c$, and kaon/proton separation up to
$p_{T}$$\sim$4.5 GeV/$c$. Together with the previously installed
Aerogel Cherenkov counter (ACC), the TOFw detector provides high
$p_{T}$ hadron identification in PHENIX. Combining the photon
yield measured in the ACC and the mass-squared from TOFw, the kaon
identification is extended to $p_{T}$$\sim 4$ GeV/$c$, while the
pion and proton identification reaches $p_{T}\sim 7$GeV/$c$.
PHENIX was also upgraded with a new reaction plane detector (RxNP)
which covers the rapidity region $1.0<|\eta|< 2.8$ with best
resolution around $74\%$ for $v_{2}$ measurements. Since the RxNP
is installed away from mid-rapidity, the non-flow effects from jet
correlation are relatively small.

\begin{figure}
\begin{center}\vspace{-0.2cm}
\includegraphics[width=9cm]{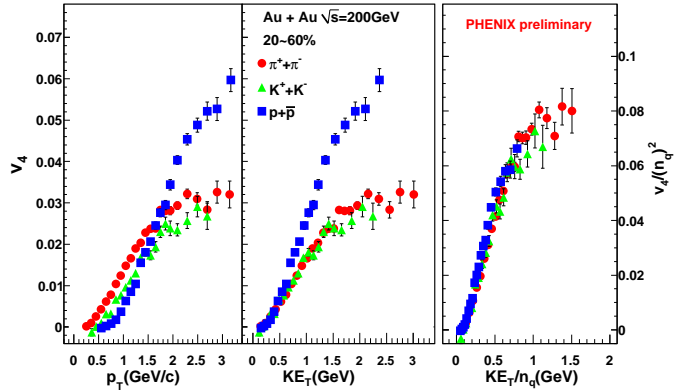}
\end{center}
\caption{The $v_{4}$ of pions, kaons and protons for $20-60\%$
Au+Au collisions at $\sqrt{s_{NN}}= 200$ GeV}
\end{figure}

\begin{figure}
\begin{center}\vspace{-0.2cm}
\includegraphics[width=7cm]{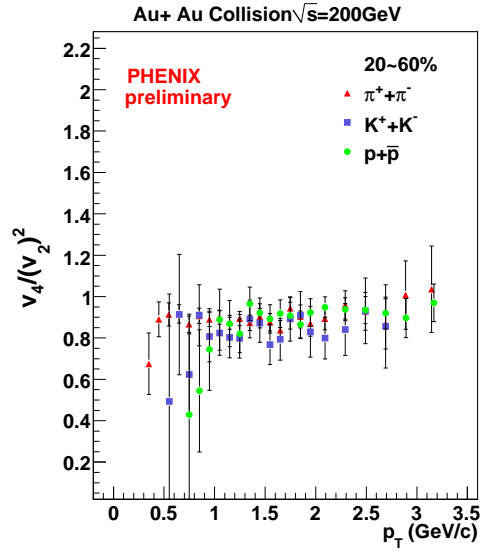}
\end{center}
\caption{The $v_{4}/v_{2}^{2}$  for pions, kaons and protons  as a
function of $p_{T}$ in the $20-60\%$ centrality class in
$\sqrt{s_{NN}}= 200$ GeV Au+Au collisions.}
\end{figure}

\begin{figure}
\begin{center}\vspace{-0.2cm}
\includegraphics[width=10cm]{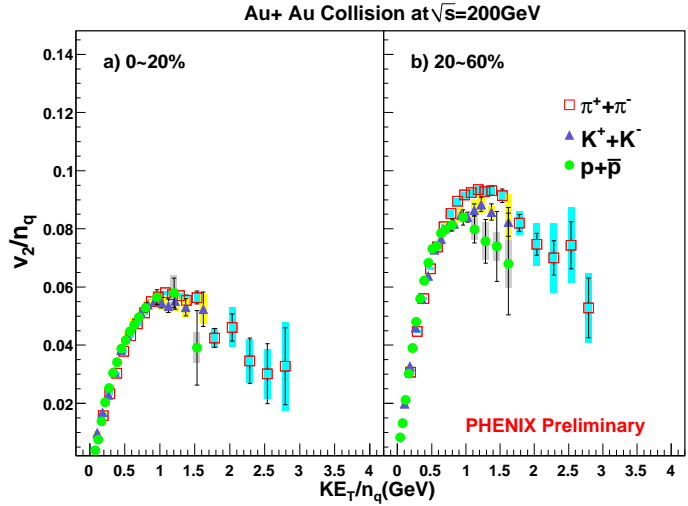}
\end{center}
\caption{Constituent quark scaling of elliptic flow, $v_{2}$ for
pions, kaons and protons as a function of transverse kinetic
energy per quark (KE$_{T}$) measured in two centrality
classes:(left)$0-20\%$,  and (right) $20-60\%$.}
\end{figure}

\section{Results}
Figure 1 shows the $v_{4}$ of pions, kaons and protons in the
$20-60\%$ centrality bin in 200 GeV Au+Au collisions. In the left
plot, the $v_{4}$ is shown as a function of $p_{T}$. A clear mass
ordering is observed for pions, kaons and protons, which is
consistent with hydrodynamics behavior which has been previously
observed for the elliptic flow. In the middle plot, the $v_{4}$
measurements are presented as a function of transverse kinetic
energy $KE_{T}$ =$m_{T}$-$m_{0}$. In this unit, the mass ordering
disappears at low $KE_{T}$ which is consistent with hydrodynamic
predictions. For $KE_{T}$ greater than about 0.5 GeV, kaons and
pions show much less $v_{4}$ than protons. A universal behavior
for baryons and mesons is observed when $KE_{T}$ is divided by the
$n_{q}$ (number of constituent quarks) and the $v_{4}$ values are
divided by the $n_{q}^{2}$. This universal behavior has also been
observed in the measurements of $v_{2}$ for identified
hadrons{\cite{flow}}. The results presented here further
strengthen the conclusion that partonic flow has been built up in
the early stages of the heavy-ion collisions at RHIC.

Figure 2 shows the $v_{4}/v_{2}^{2}$ ratio for pions, kaons and
protons as a function of $p_{T}$ in the $20-60\%$ centrality bin.
This ratio is flat with $p_{T}$ in the measured range and is
independent of the particle species within errors. We analyze the
results in terms of a simple coalescence model:
\numparts
\begin{eqnarray}
\frac{v_{4,m}(2p_{T})}{v^{2}_{2,m}(2p_{T})}=\alpha(\frac{1}{4}+\frac{1}{2}
\frac{v_{4,q}(p_{T})}{v^{2}_{2,q}(p_{T})})\\
\frac{v_{4,b}(3p_{T})}{v^{2}_{2,b}(3p_{T})}=\alpha(\frac{1}{3}+\frac{1}{3}
\frac{v_{4,q}(p_{T})}{v^{2}_{2,q}(p_{T})})
\end{eqnarray}
\endnumparts
where $v_{4,m}(p_{T})$, $v_{4,b}(p_{T})$ and $v_{4,q}(p_{T})$
represent the meson, baryon and quark $v_{4}$ respectively, and
$v_{2,m}(p_{T})$, $v_{2,b}(p_{T})$ and $v_{2,q}(p_{T})$ represent
the meson, baryon and quark $v_{2}$. Using the measured
$v_{4}/v_{2}^{2}$ ratio around 0.9 for both baryons and mesons,
from equations (1a) and (1b), we obtain that the parton
$v_{4,q}/v_{2,q}^{2}$ ratio is around 0.5 and the parameter
$\alpha$ is around 1.8. This result indicates that a thermalized
partonic liquid has been produced at RHIC.

Using the high $p_{T}$ elliptic flow results, we can study the
limits of applicability of the hydrodynamic description. Figure 3
shows the $v_{2}$ of pions, kaons and protons as a function of
$KE_{T}$ in two centrality bins. Both $v_{2}$ and $KE_{T}$ have
been divided by the $n_{q}$. The left plot is the result in the
$0-20\%$ centrality bin, and the right plot is the result in the
$20-60\%$ centrality bin. In $20-60\%$ collisions, the NCQ scaling
begins to break as the $KE_{T}$/$n_{q}$ exceeds $\sim$1 GeV. This
indicates that the origin of the $v_{2}$ is based in hydrodynamics
collective flow and parton recombination in the low $KE_{T}$
region, but above $KE_{T}/n_{q}\sim 1$ GeV, the contribution from
parton energy loss become increasingly important.

\section{Conclusions}
The measurements of pion, kaon and proton $v_{2}$ and $v_{4}$ have
been extended up to a $p_{T}$ of 6 GeV/c and 4 GeV/c respectively
by PHENIX. The NCQ scaling has been tested for $v_{4}$ and been
found to hold for $KE_{T}$/$n_{q}$ up to 1 GeV, indicating that
partonic flow governs the bulk dynamics in heavy-ion collisions at
RHIC. The mesons and baryons have a similar ratio of
$v_{4}/v_{2}^{2}$, which is consistent with expectations for a
thermalized partonic system in which hadrons at the intermediate
$p_{T}$ region are produced by parton recombination. The $v_{2}$
measurement shows that the NCQ scaling begins to break for
$KE_{T}$/$n_{q}$ above 1 GeV in the $20-60\%$ centrality class,
which suggests that hard-scattering may be the dominant production
mechanism for both baryons and mesons in this $KE_{T}$/$n_{q}$
range and thus parton energy loss effects play a significant role
in generating the azimuthal anisotropy in particle emission.


\section{References}
\begin{thebibliography}{99}
\bibitem{whitepaper} K. Adcox et al., Nucl. Phys. A 757, 184 (2005).
\bibitem{photon} A. Adare et al., arXiv:0804.4168[nucl-ex].
\bibitem{flow} A. Adare et al. Phys. Rev. Lett. 99, 052301 (2007); B. I. Abelev et al., Phys. Rev. C 75, 054906
(2007).
\bibitem{thermalization} Peter Kolb, Phys. Rev. C 68, 031902 (2003);
R. S. Bhalerao et.al, Phys. Lett. B 627, 49 (2005); Borghini N and
Ollitrault, J-Y, Phys. Lett. B 642, 227 (2006); Ko. C. M, J. Phys.
G: Nucl. Part. Phys. 34 S413(20) (2007)
\bibitem{jet} R. J. M. Snellings, A. M. Poskanzer, and S. A. Voloshin,
arXiv:9904003[nucl-ex]; X. N. Wang, Phys. Rev. C 63, 054902
(2001).

\bibitem{Hwa} R. C. Hwa and C. B. Yang, arXiv:0801.2183[nucl-th].
\end{thebibliography}
\end{document}